\documentclass[usenatbib,useAMS]{mnras}

\usepackage{lmodern}

\usepackage{natbib}
\usepackage{amsmath}
\usepackage{graphicx}
\usepackage{color}
\usepackage{mathrsfs}
\usepackage{epsfig}
\usepackage{comment}
\usepackage{ulem}

\usepackage{graphicx}
\usepackage[hang,small,bf]{caption}
\usepackage[subrefformat=parens]{subcaption}
\usepackage{amssymb}
\usepackage[T1]{fontenc}

\newcommand{\msun}{{M}_{\odot}}

\newcommand{\Mc}{M_{\rm chirp}}

\newcommand{\beq}{\begin{equation}}
\newcommand{\eeq}{\end{equation}}

\defcitealias{K14}{K14}
\defcitealias{K16}{K16}
\defcitealias{Visbal_2015}{VHB15}

\usepackage{url} 

\title[Mass Ratio of BBHs Determined from LIGO/Virgo Data]
{Mass Ratio of Binary Black Holes Determined from LIGO/Virgo Data Restricted to Small False Alarm Rate}

\author[T. Kinugawa, T. Nakamura, and H. Nakano]
{
Tomoya Kinugawa$^{(1)(2)(3)}$\thanks{E-mail: kinugawa@shinshu-u.ac.jp}, Takashi Nakamura$^{(4)}$, and Hiroyuki Nakano$^{(5)}$ \\
\\
$^{1}$Faculty of Engineering, Shinshu University, 
Nagano 380-8553, Japan\\
$^{2}$Research Center for Aerospace System, Shinshu University, 
Nagano 380-8553, Japan\\
$^{3}$Research Center for the Early Universe, Graduate School of Science, University of Tokyo, 
Tokyo 113-0033, Japan\\
$^{4}$Department of Physics, Graduate School of Science, Kyoto University,
Kyoto 606-8502, Japan\\
$^{5}$Faculty of Law, Ryukoku University, Kyoto 612-8577, Japan
}

\begin{document}

\date{\today}
\maketitle
\begin{abstract}
We focus on gravitational-wave events of binary black-hole mergers up to the third observing run with the minimum false alarm rate smaller than $10^{-5}\,{\rm yr}^{-1}$. These events tell us that the mass ratio of two black holes follows $m_2/m_1=0.723$ with the chance probability of 0.00301\% {for the chirp mass $\Mc > 18\,M_{\odot}$}. We show that the relation of $m_2/m_1=0.723$ is consistent with the binaries originated from population III stars which are the first stars in the universe. On the other hand, it is found for $\Mc < 18 M_{\odot}$ that the mass ratio follows $m_2/m_1=0.601$ with the chance probability of 0.117\% if we ignore GW190412 with $m_2/m_1\sim 0.32$. This suggests a different origin from that for $\Mc > 18 M_{\odot}$.
\end{abstract}


\section{Introduction}

Possible origin of massive binary black holes (BBHs) with total mass $\sim 65\,\msun$ like GW150914 which is the world's first observation of gravitational waves (GWs), is Population (Pop) III stars. 
Theoretically, GW events like GW150914 were predicted by~\cite{Kinugawa2014} before the discovery of GW150914 by \cite{Abbott_PRL_2016}. 
After that, various compact object binaries have been observed by LIGO/Virgo GW detectors.

Many of them are BBHs (see, e.g., the third Gravitational-wave Transient Catalog (GWTC-3) of~\cite{2021arXiv211103606T}).
{\cite{2021arXiv211103634T} found, e.g., the merger rate of $16$--$130\,{\rm yr}^{-1}{\rm Gpc}^{-3}$ and the substructure in the chirp mass distribution with peak around $8\,M_{\odot}$, weak structure around $15\,M_{\odot}$, and peak around $30\,M_{\odot}$ for the BBH events with the minimum false alarm rate (FAR), ${\rm \bf FAR}_{\rm min} < 1\,{\rm yr}^{-1}$.
Here, the chirp mass $M_{\rm chirp}$ is defined as
\begin{equation}
M_{\rm chirp} = \frac{(m_1 m_2)^{3/5}}{(m_1+m_2)^{1/5}} \,,
\end{equation}
where $m_1$ and $m_2$ ($ \leq m_1$) denote the primary and secondary mass, respectively.
${\rm \bf FAR}_{\rm min}$ is evaluated by various pipelines used in the GW data analysis.}
As for the distribution of mass ratio $q$, 
\begin{equation}
q =\frac{m_2}{m_1} \,,
\end{equation}
a power law was treated to model it (see, e.g.,~\cite{2021arXiv210714239M} for rates of compact object coalescences, and references therein).

In the following, we focus only on BBHs, and especially the masses~\footnote{There are also studies on the spins, for example, see \cite{2022arXiv220702924F} for limits on hierarchical BH mergers from an effective inspiral spin parameter. To extract more detailed information of spins, we will require multiband GW observations~\citep{Isoyama:2018rjb} with a decihertz GW detector, B-DECIGO~\citep{Nakamura_2016}.}.
Although the mass ratio of binaries is estimated less accurately than the chirp mass, we can find several studies on the mass ratio related to~\cite{2021arXiv211103634T}.
\cite{2021arXiv211113991T} showed that there is no prominent dependence either on the chirp mass or the aligned spin in the mass ratio distribution from the 69 GW events with ${\rm \bf FAR}_{\rm min} < 1\,{\rm yr}^{-1}$.
Here, it should be noted that the estimation of spins is more difficult than the mass ratio.
In our galaxy, \cite{2021arXiv211113704W} give a prediction of the mass ratio distribution with a peak at $q \approx 0.4$ for BBHs observed by a 4\,yr LISA observation~\citep{2017arXiv170200786A} in their fiducial model.
In simulations for hierarchical triples from low-mass young star clusters, \cite{2022MNRAS.511.1362T} found $q \approx 0.3$ which is lower than that from binaries (see, e.g., \cite{2021MNRAS.507.3612R}).
\cite{2021arXiv211205763B} showed that at least 95\% of BBH mergers detectable by LIGO/Virgo/KAGRA (LVK) detector network at design sensitivity have $q \gtrsim 0.25$ in their 560 models.
\cite{2021arXiv211210786S} presented orbital properties (masses, mass ratio, eccentricity, orbital separation etc.) of surviving systems after a BBH has formed in the inner binary and those of BBHs which are formed from an isolated binary population for metallicities, $Z=0.01\,Z_{\odot}$ and $Z_{\odot}$, where $Z_{\odot}$ is the solar metallicity, obtained by using a new triple stellar evolution code.
Using the 69 GW events with ${\rm \bf FAR}_{\rm min} < 1\,{\rm yr}^{-1}$, \cite{2022arXiv220101905L} found that the observed BBHs have a much stronger preference for equal mass binaries in their parameterized primary-mass distribution models.
Using direct $N$-body simulations for star cluster models, \cite{2022arXiv220208924C} found that the distributions of the mass ratio have median values in the range of $0.8$--$0.9$ (except for one model).
\cite{2022arXiv220303651M} have prepared a deep-learning pipeline to constrain properties of hierarchical black-hole mergers.
In the field of galaxies, \cite{2022arXiv220316544S} discussed BBHs starting from hierarchical triple population and isolated binary population, and found that the observed lower mass ratio ($q \lesssim 0.5$) BBHs can be explained by the contribution from the outer binary channel of the triple population.
In mass-ratio reversal systems where the second BH to form in the binary are more massive than the first BH, \cite{2022arXiv220501693B} found that BBHs with $M_{\rm chirp} \gtrsim 10\,M_{\odot}$ and $q \gtrsim 0.6$ are dominant in the GW observation (see also \cite{2022arXiv220512329M}). 
\cite{2022arXiv220613842B} introduced super-Eddington accretion into a population synthesis code.
\cite{2022arXiv220801081A} discussed a scenario of merging BBHs which are formed dynamically in globular clusters, and found that the observed events shown in GWTC-3 cannot be explained in the above scenario (see also~\cite{2022arXiv220905766M}).
In \cite{2022arXiv220905959F,2022arXiv220906196E}, the mass ratio has been discussed by treating a varying equation of state at the QCD epoch in primordial BH formation scenario (see also \cite{2023arXiv230603903C} for a recent review on primordial BHs).
\cite{2022arXiv221012834E} have suggested some possible plateaus at several mass ratios in the distribution by using a data-driven, non-parametric model.
\cite{2023arXiv230306081O} have discussed merging compact binaries with highly asymmetric mass ratios, and found that 
a natal kick of $\sim 250 \,{\rm km/s}$ is required for systems with $q \lesssim 0.1$ to merge.
\cite{2023arXiv230315511C} discussed BBHs from Pop II ($Z = 10^{-4}$) and III ($Z = 10^{-11}$) stars in their models, and found that the mass ratios for Pop II BBHs are almost $q \sim 1$ and the peak of mass ratios for Pop III BBHs is $q=0.8$ -- $0.9$ in most of their models.
In \cite{2023arXiv230315515S}, the redshift dependence of mass ratio of Pop III BBHs was presented.
In their models, Pop III BBHs merging at low redshift ($z \leq 4$) have low mass ratios, $q \approx 0.5$ -- $0.7$ (the median values), while typically $q \sim 0.9$ at high redshift.
For star clusters, \cite{2023arXiv230704807A} have investigated dependence of the host cluster structure and the physics of massive star evolution by using their database which has 19 direct N-body models to discuss various types of compact binary mergers~\citep{2023arXiv230704805A}, and found $q > 0.6$ and $m_1/\msun = 5$-- $40$ in primordial binary systems, and $q=0.6$--$1$ for $m_1 < 15\,\msun$, $q>0.9$ for $15\,\msun < m_1 < 40.5\,\msun$ and $q>0.7$ for $40.5\,\msun < m_1$ in dynamical mergers.

\section{Analysis of data with small false alarm rate}

The GW events with ${\rm \bf FAR}_{\rm min} < 1 \times 10^{-5}\,{\rm yr}^{-1}$ in~\cite{2021arXiv211103634T} are summarized in Table~\ref{tab:events} {(note that because there are some updates during this work, we use the latest ones from the online GWTC~\citep{online_GWTC})}.
{The reason for this restriction of FAR is that we treat more accurate tendency of the events.} 
Here, we show the event name, primary mass $m_1$, secondary mass $m_2$, chirp mass $M_{\rm chirp}$, luminosity distance $D_{\rm L}$, and mass ratio $q$.
BNS, NS-BH, MGCO-BH, and BBH in the column of ``Binary type'' mean binary neutron star, neutron star-black hole binary, mass-gap compact object~\footnote{The mass of MGCOs lies in $2$--$5\,M_{\odot}$. See~\cite{2021PTEP.2021b1E01K} for details.}-black hole binary, and binary black hole, respectively.

\begin{table*}
\caption{Event name (the YYMMDD\_hhmmss format), primary mass $m_1$, secondary mass $m_2$, chirp mass $M_{\rm chirp}$ in unit of the solar mass, $M_{\odot}$, and luminosity distance $D_{\rm L}$ [Mpc] from~\citet{online_GWTC}.
These are expressed by the median and 90\%-symmetric credible interval.
The mass ratio $q$ is evaluated by using the median values
of $m_1$ and $m_2$.
Binary type shows binary neutron star (BNS), neutron star-black hole binary (NS-BH), mass-gap compact object-black hole binary (MGCO-BH), or binary black hole (BBH).
Here, we focus only on events with ${\rm \bf FAR}_{\rm min} < 1 \times 10^{-5}\,{\rm yr}^{-1}$ where ${\rm \bf FAR}_{\rm min}$ means the minimum FAR evaluated by various GW data analyses.
These events have the probability of astrophysical (signal) origin, $p_{\rm astro} > 0.99$.
The data are sorted by $M_{\rm chirp}$.
The first 2 events are BNSs and NS-BH binaries.
The next 14 events are BBHs and a MGCO-BH, and have $M_{\rm chirp} < 18\,M_{\odot}$, and the final 20 events are BBHs and have $M_{\rm chirp} > 18\,M_{\odot}$.}
\label{tab:events}
\begin{center}
\renewcommand{\arraystretch}{1.2}
\begin{tabular}{ccccccc}
\hline
Event name & $m_1$ & $m_2$ & $M_{\rm chirp}$ & $D_{\rm L}$ & $q$ & Binary type \\
\hline
GW170817 & $1.46_{-0.1}^{+0.12}$ & $1.27_{-0.09}^{+0.09}$ & $1.186_{-0.001}^{+0.001}$ & $40_{-15}^{+7}$ & 0.87 & BNS
\\
GW200115\_042309 & $5.9_{-2.5}^{2}$ & $1.44_{-0.29}^{+0.85}$ & $2.43_{-0.07}^{+0.05}$ & $290_{-100}^{+150}$ & 0.24 & NS-BH \\
\hline
GW190924\_021846 & $8.8_{-1.8}^{+4.3}$ & $5.1_{-1.5}^{+1.2}$ & $5.8_{-0.2}^{+0.2}$ & $550_{-220}^{+220}$ & 0.58 & BBH \\
GW190814\_211039 & $23.3_{-1.4}^{+1.4}$ & $2.6_{-0.1}^{+0.1}$ & $6.11_{-0.05}^{+0.06}$ & $230_{-50}^{+40}$ & 0.11 & MGCO-BH \\
GW191129\_134029 & $10.7_{-2.1}^{+4.1}$ & $6.7_{-1.7}^{+1.5}$ & $7.31_{-0.28}^{+0.43}$ & $790_{-330}^{+260}$ & 0.63 & BBH \\
GW200202\_154313 & $10.1_{-1.4}^{+3.5}$ & $7.3_{-1.7}^{+1.1}$ & $7.49_{-0.2}^{+0.24}$ & $410_{-160}^{+150}$ & 0.72 & BBH \\
GW170608 & $11_{-1.7}^{+5.5}$ & $7.6_{-2.2}^{+1.4}$ & $7.9_{-0.2}^{+0.2}$ & $320_{-110}^{+120}$ & 0.69 & BBH \\
GW191216\_213338 & $12.1_{-2.3}^{+4.6}$ & $7.7_{-1.9}^{+1.6}$ & $8.33_{-0.19}^{+0.22}$ & $340_{-130}^{+120}$ & 0.64 & BBH \\
GW190707\_093326 & $12.1_{-2}^{+2.6}$ & $7.9_{-1.3}^{+1.6}$ & $8.4_{-0.4}^{+0.6}$ & $850_{-400}^{+340}$ & 0.65 & BBH \\
GW191204\_171526 & $11.9_{-1.8}^{+3.3}$ & $8.2_{-1.6}^{+1.4}$ & $8.55_{-0.27}^{+0.38}$ & $650_{-250}^{+190}$ & 0.69 & BBH \\
GW190728\_064510 & $12.5_{-2.3}^{+6.9}$ & $8_{-2.6}^{+1.7}$ & $8.6_{-0.3}^{+0.6}$ & $880_{-380}^{+260}$ & 0.64 & BBH \\
GW200316\_215756 & $13.1_{-2.9}^{+10.2}$ & $7.8_{-2.9}^{+1.9}$ & $8.75_{-0.55}^{+0.62}$ & $1120_{-440}^{+470}$ & 0.60 & BBH \\
GW151226 & $13.7_{-3.2}^{+8.8}$ & $7.7_{-2.5}^{+2.2}$ & $8.9_{-0.3}^{+0.3}$ & $450_{-190}^{+180}$ & 0.56 & BBH \\
GW190720\_000836 & $14.2_{-3.3}^{+5.6}$ & $7.5_{-1.8}^{+2.2}$ & $9_{-0.8}^{+0.4}$ & $770_{-260}^{+650}$ & 0.53 & BBH \\
GW190412\_053044 & $27.7_{-6}^{+6}$ & $9_{-1.4}^{+2}$ & $13.3_{-0.5}^{+0.5}$ & $720_{-220}^{+240}$ & 0.32 & BBH \\
GW190512\_180714 & $23.2_{-5.6}^{+5.6}$ & $12.5_{-2.6}^{+3.5}$ & $14.6_{-0.9}^{+1.4}$ & $1460_{-590}^{+510}$ & 0.54 & BBH \\
\hline
GW191215\_223052 & $24.9_{-4.1}^{+7.1}$ & $18.1_{-4.1}^{+3.8}$ & $18.4_{-1.7}^{+2.2}$ & $1930_{-860}^{+890}$ & 0.73 & BBH \\
GW190408\_181802 & $24.8_{-3.5}^{+5.4}$ & $18.5_{-4}^{+3.3}$ & $18.5_{-1.2}^{+1.9}$ & $1540_{-620}^{+440}$ & 0.75 & BBH \\
GW170104 & $30.8_{-5.6}^{+7.3}$ & $20_{-4.6}^{+4.9}$ & $21.4_{-1.8}^{+2.2}$ & $990_{-430}^{+440}$ & 0.65 & BBH \\
GW170814 & $30.6_{-3}^{+5.6}$ & $25.2_{-4}^{+2.8}$ & $24.1_{-1.1}^{+1.4}$ & $600_{-220}^{+150}$ & 0.82 & BBH \\
GW190915\_235702 & $32.6_{-4.9}^{+8.8}$ & $24.5_{-5.8}^{+4.9}$ & $24.4_{-2.3}^{+3}$ & $1750_{-650}^{+710}$ & 0.75 & BBH \\
GW190828\_063405 & $31.9_{-4.1}^{+5.4}$ & $25.8_{-5.3}^{+4.9}$ & $24.6_{-2}^{+3.6}$ & $2070_{-920}^{+650}$ & 0.81 & BBH \\
GW170809 & $35_{-5.9}^{+8.3}$ & $23.8_{-5.2}^{+5.1}$ & $24.9_{-1.7}^{+2.1}$ & $1030_{-390}^{+320}$ & 0.68 & BBH \\
GW190630\_185205 & $35.1_{-5.5}^{+6.5}$ & $24_{-5.2}^{+5.5}$ & $25.1_{-2.1}^{+2.2}$ & $870_{-360}^{+530}$ & 0.68 & BBH \\
GW200311\_115853 & $34.2_{-3.8}^{+6.4}$ & $27.7_{-5.9}^{+4.1}$ & $26.6_{-2}^{+2.4}$ & $1170_{-400}^{+280}$ & 0.81 & BBH \\
GW200129\_065458 & $34.5_{-3.2}^{+9.9}$ & $28.9_{-9.3}^{+3.4}$ & $27.2_{-2.3}^{+2.1}$ & $900_{-380}^{+290}$ & 0.84 & BBH \\
GW200112\_155838 & $35.6_{-4.5}^{+6.7}$ & $28.3_{-5.9}^{+4.4}$ & $27.4_{-2.1}^{+2.6}$ & $1250_{-460}^{+430}$ & 0.79 & BBH \\
GW150914 & $35.6_{-3.1}^{+4.7}$ & $30.6_{-4.4}^{+3}$ & $28.6_{-1.5}^{+1.7}$ & $440_{-170}^{+150}$ & 0.86 & BBH \\
GW170823 & $39.5_{-6.7}^{+11.2}$ & $29_{-7.8}^{+6.7}$ & $29.2_{-3.6}^{+4.6}$ & $1940_{-900}^{+970}$ & 0.73 & BBH \\
GW190503\_185404 & $41.3_{-7.7}^{+10.3}$ & $28.3_{-9.2}^{+7.5}$ & $29.3_{-4.4}^{+4.5}$ & $1520_{-600}^{+630}$ & 0.69 & BBH \\
GW190727\_060333 & $38.9_{-6}^{+8.9}$ & $30.2_{-8.3}^{+6.5}$ & $29.4_{-3.7}^{+4.6}$ & $3070_{-1230}^{+1300}$ & 0.78 & BBH \\
GW200224\_222234 & $40_{-4.5}^{+6.9}$ & $32.5_{-7.2}^{+5}$ & $31.1_{-2.6}^{+3.2}$ & $1710_{-640}^{+490}$ & 0.81 & BBH \\
GW190521\_074359 & $43.4_{-5.5}^{+5.8}$ & $33.4_{-6.8}^{+5.2}$ & $32.8_{-2.8}^{+3.2}$ & $1080_{-530}^{+580}$ & 0.77 & BBH \\
GW191222\_033537 & $45.1_{-8}^{+10.9}$ & $34.7_{-10.5}^{+9.3}$ & $33.8_{-5}^{+7.1}$ & $3000_{-1700}^{+1700}$ & 0.77 & BBH \\
GW190519\_153544 & $65.1_{-11}^{+10.8}$ & $40.8_{-12.7}^{+11.5}$ & $44.3_{-7.5}^{+6.8}$ & $2600_{-960}^{+1720}$ & 0.63 & BBH \\
GW190602\_175927 & $71.8_{-14.6}^{+18.1}$ & $44.8_{-19.6}^{+15.5}$ & $48_{-9.7}^{+9.5}$ & $2840_{-1280}^{+1930}$ & 0.62 & BBH \\
\hline
\end{tabular}
\renewcommand{\arraystretch}{1.0}
\end{center}
\end{table*}

The study by~\cite{2020MNRAS.498.3946K} is based on our old code developed by~\cite{Kinugawa2014} for population synthesis simulations of the evolution of Pop III stars. There are, however, at least two big differences between the two papers. The first one is that the star formation rate (SFR) of Pop III stars is a factor 3 decreased due to the new observational data of {CMB (Cosmic Microwave Background)}. The second one is the number of simulated models. In~\cite{Kinugawa2014}, only two models with different initial mass function (IMF) are simulated, while in~\cite{2020MNRAS.498.3946K} seven models with different IMF, initial mass ratio, initial separation, initial eccentricity, mass transfer rate, accretion fraction ($\beta$), common envelope parameter $\alpha \lambda$ and tidal coefficient factor ($E$) are simulated.

\begin{figure}
  \begin{center}
    \includegraphics[width=\hsize]{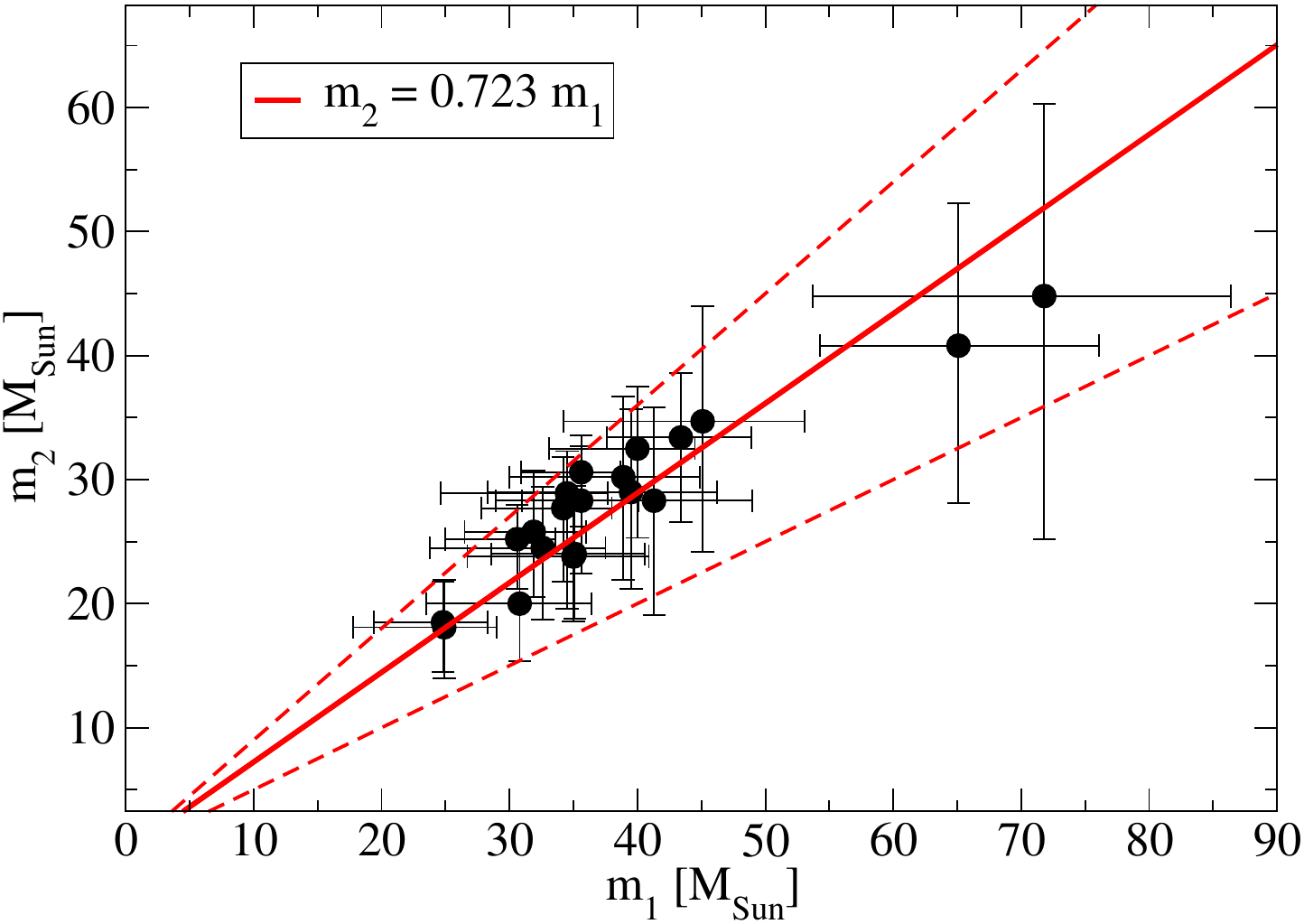}
  \end{center}
  \caption{20 BBH events with $M_{\rm chirp} > 18\,M_{\odot}$ in Table~\ref{tab:events}. We present the median and 90\%-symmetric credible interval for $m_1$ and $m_2$. The fitting (the solid red line) gives $m_2=0.723\, m_1$, and the correlation coefficient is $0.933$ with the chance probability of 0.00301\%. As references, we show $m_2=0.5\, m_1$ and $m_2=0.9\, m_1$ as the dashed red lines.}
  \label{fig:m1_m2_high}
\end{figure}

In our previous study~\citep{2021MNRAS.504L..28K}, we found a very simple relation, $m_2\simeq 0.7\,m_1$~\footnote{\cite{2022arXiv220205861B} also mentioned this relation for nearly all BBH GW events.} for BBHs with $M_{\rm chirp} \gtrsim 20\,M_{\odot}$ summarized in GWTC-2~\citep{2021PhRvX..11b1053A}, and that this relation is consistent with the mass distribution in our 6 population synthesis simulations of Pop III stars in Fig.~4 of~\cite{2021MNRAS.504L..28K}. Therefore, first, we focus only on 20 BBH events with $M_{\rm chirp} > 18\,M_{\odot}$.
This is because there exists a small gap in the chirp mass between GW190512\_180714 and GW191215\_223052. 
Figure~\ref{fig:m1_m2_high} shows $m_1$ and $m_2$ of these events. 
Assuming $m_2=0$ at $m_1=0$, the linear fitting function is obtained as 
\begin{equation}
m_2=0.723\, m_1 \,, 
\end{equation}
and the correlation coefficient is $0.933$ with the chance probability of 0.00301\%.

\begin{figure}
  \begin{center}
    \includegraphics[width=\hsize]{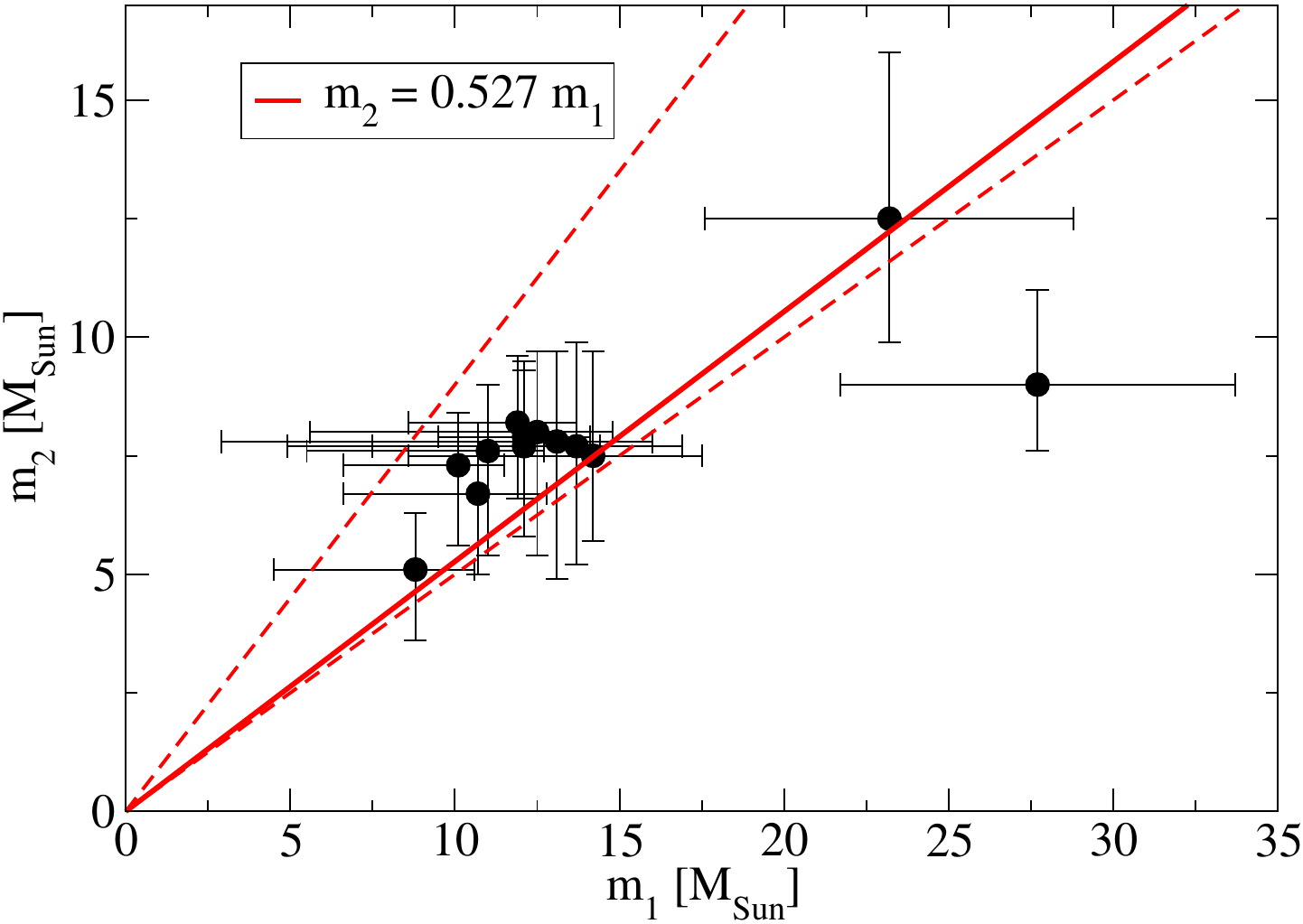}
  \end{center}
  \caption{13 BBH events with $M_{\rm chirp} < 18\,M_{\odot}$ in Table~\ref{tab:events}. We present the median and 90\%-symmetric credible interval for $m_1$ and $m_2$. The fitting (the solid red line) gives $m_2=0.527\, m_1$, and the correlation coefficient is $0.741$. As references, we show $m_2=0.5\, m_1$ and $m_2=0.9\, m_1$ as the dashed red lines.}
  \label{fig:m1_m2_low}
\end{figure}

Next, we treat the remaining BBH events. Figure~\ref{fig:m1_m2_low} shows $m_1$ and $m_2$ of 13 BBH events with $M_{\rm chirp} < 18\,M_{\odot}$ given in Table~\ref{tab:events}.
Here, we have ignored GW190814\_211039 which has a MGCO in the binary.
Assuming $m_2=0$ at $m_1=0$, the linear fitting function is obtained as $m_2=0.527\, m_1$, and the correlation coefficient is $0.741$.
The correlation between the data and this fitting function
is not so high.

\begin{figure}
  \begin{center}
    \includegraphics[width=\hsize]{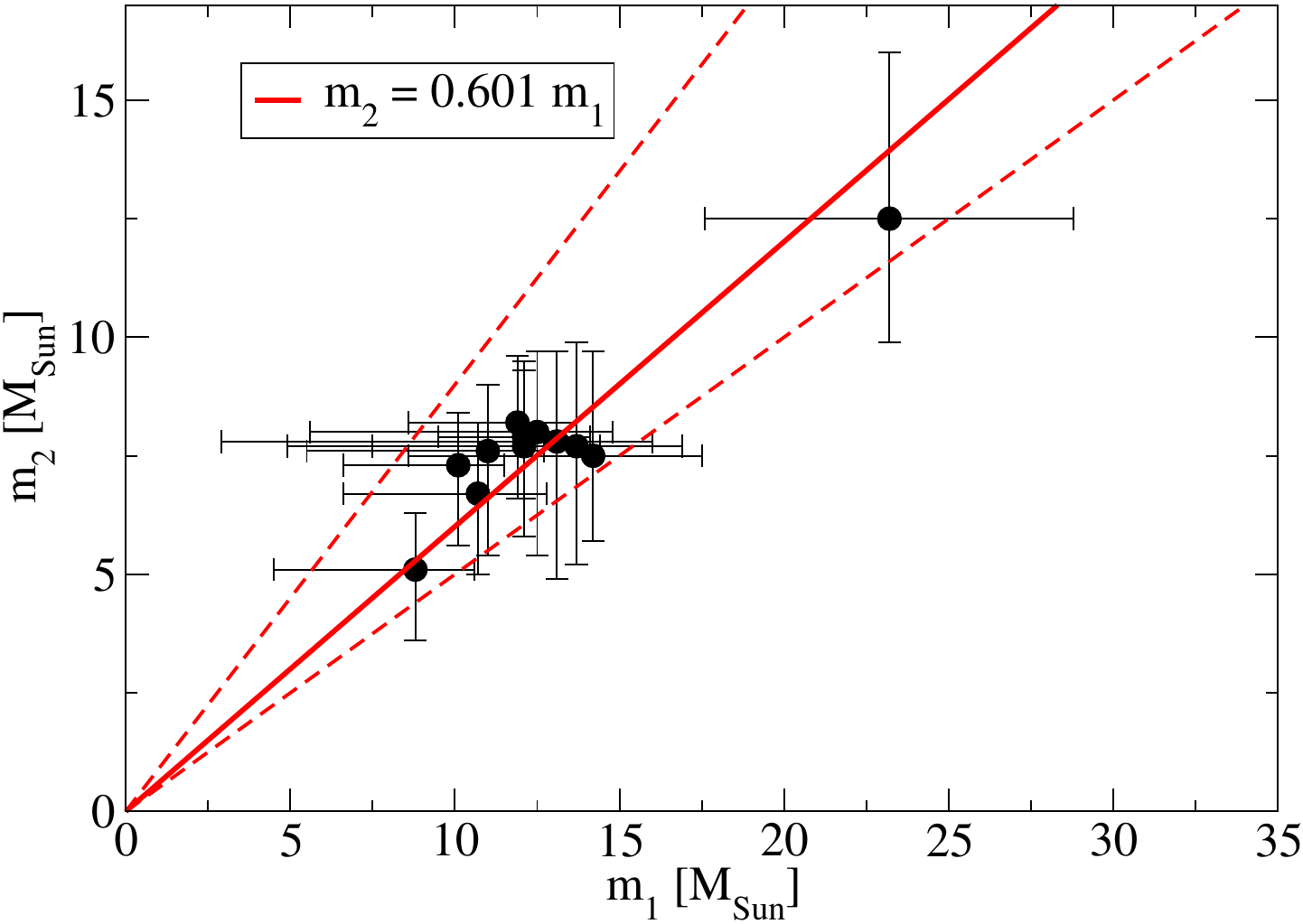}
  \end{center}
  \caption{12 BBH events with $M_{\rm chirp} < 18\,M_{\odot}$ in Table~\ref{tab:events}. We present the median and 90\%-symmetric credible interval for $m_1$ and $m_2$. Here, we have ignored a BBH event, GW190412\_053044. The fitting (the solid red line) gives $m_2=0.601\, m_1$, and the correlation coefficient is $0.937$ with the chance probability of 0.117\%. As references, we show $m_2=0.5\, m_1$ and $m_2=0.9\, m_1$ as the dashed red lines.}
  \label{fig:m1_m2_low_c}
\end{figure}

Here, we note that GW190412\_053044 has large unequal component masses~\citep{2020PhRvD.102d3015A}.
As an alternative interpretation of masses for this GW event, we may have $q = 0.31^{+0.05}_{-0.04}$ from a prior with a non-spinning primary and a rapidly spinning secondary \citep{2020ApJ...895L..28M} (see also \cite{2020MNRAS.498.3946K}).
Also, \cite{2020ApJ...899L..17Z} found $q \lesssim 0.57$ in the 99\% credible level from various models.
Interestingly, although the GW data did not prefer Model G with a prior assumption, $\chi_1=\chi_2=0$ in the above paper, this model gave $q \approx 0.55$.
\cite{2023arXiv230611088A} have also discussed the most likely formation channel.
When we ignore this GW190412\_053044 event in the analysis of mass ratio, the fitting is improved as Fig.~\ref{fig:m1_m2_low_c}.
Assuming $m_2=0$ at $m_1=0$, the linear fitting function is obtained as
\begin{equation}
m_2=0.601\, m_1 \,, 
\end{equation}
and the correlation coefficient becomes $0.937$ with the chance probability of 0.117\%.
To argue the origin of BBH events for $M_{\rm chirp} < 18\,M_{\odot}$, we need more examples of BBHs similar to GW190412\_053044 in the near future by the O4 Observing run~\citep{2020LRR....23....3A}.

In \cite{2020MNRAS.498.3946K, 2021MNRAS.504L..28K}, we performed $10^6$ Pop III binary evolution by using 12 different models with initial conditions of mass function, mass ratio, separation, and eccentricity as well as physical models (see Tables 2 and 3 of~\cite{2020MNRAS.498.3946K} and Section 2.2 of~\cite{2021MNRAS.504L..28K} for details of each model). 
These Pop III binary population synthesis calculations are based on the Pop III stellar evolution model of \cite{Marigo_2001}. The fiducial model of \cite{2020MNRAS.498.3946K} assumes the flat IMF from $10\,\msun$ to $150\,\msun$, the flat mass ratio distribution from $10\,\msun/M_1$ to 1, the logflat separation distribution, and the eccentricity distribution proportional to $e$ for determination of initial Pop III binary conditions. In regards to binary evolution, the fiducial model of \cite{2020MNRAS.498.3946K} uses the common envelope parameter $\alpha\lambda=1$, the conservative mass transfer, and no BH natal kick. The al01 model of \cite{2020MNRAS.498.3946K} uses the same initial distributions and binary parameters of the fiducial model, except for the common envelope parameter $\alpha\lambda=0.1$.
The MT05 model of \cite{2020MNRAS.498.3946K} also uses the same initial distributions and binary parameters except for the unconservative mass transfer which loses half of accreted mass.
The M100 model of \cite{2020MNRAS.498.3946K} has modified the initial primary star mass range from $10\,\msun$ to $100\,\msun$ compared to the fiducial model. 
The K14 model of \cite{2020MNRAS.498.3946K} is based on our first Pop III calculation \citep{Kinugawa2014}.
Although this model uses the same initial distribution and binary parameters as the M100 model, the main difference between the two models is the treatment of the mass transfer rate in the stable Roche lobe overflow shown from Eq.~(4) to Eq.~(7) of~\cite{2020MNRAS.498.3946K} in details.
The FS1 and FS2 models of \cite{2020MNRAS.498.3946K} use the initial distribution and the binary parameters of FS1 and FS2 models in \cite{Belczynski_2017}, but the Pop III stellar evolution and equations of binary evolution are the same as the fiducial model of \cite{2020MNRAS.498.3946K}.
The flat model of \cite{2021MNRAS.504L..28K} is same as the M100 model of \cite{2020MNRAS.498.3946K}.
The M-05, M-1, M-15, M-2, and Sal models of \cite{2021MNRAS.504L..28K} use the $f(M_1)\propto M_1^{-0.5}$, $M_1^{-1.0}$, $M_1^{-1.5}$, $M_1^{-2}$ and Salpeter IMF, respectively.
These models use the same initial conditions and binary parameters of the flat model of \cite{2021MNRAS.504L..28K} (M100 model of \cite{2020MNRAS.498.3946K}) except for IMF.

\begin{figure}
  \begin{center}
    \includegraphics[width=\hsize]{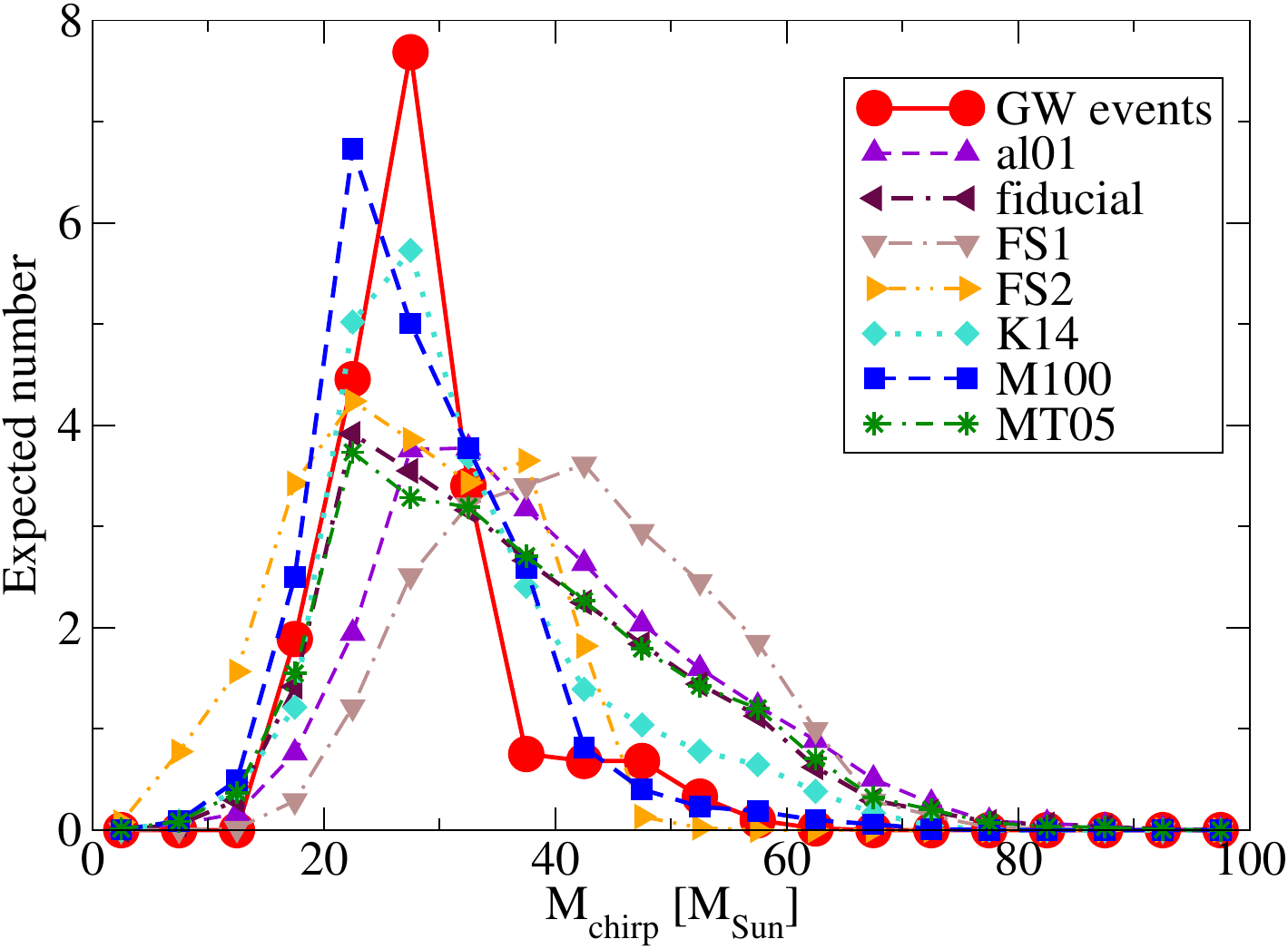}
  \end{center}
  \caption{Chirp-mass distribution of 20 BBH events with $\Mc > 18\,M_{\odot}$ for \citet{2020MNRAS.498.3946K}. The ordinate shows the expected number of BBH for given range of $\Mc$ so that the total number should be 20. The filled (red) circles show the observed 20 GW events, and the filled (purple) triangles, filled (copper) triangles, filled (mustard) triangles, filled (khaki) triangles, filled (light blue) diamonds, filled (blue) squares, and  (green) asterisks denote the results of the al01, fiducial, FS1, FS2, K14, M100 and MT05 models by assuming that the total observable number is 20.}
  \label{fig:Mc_dist20}
\end{figure}

\begin{figure}
  \begin{center}
    \includegraphics[width=\hsize]{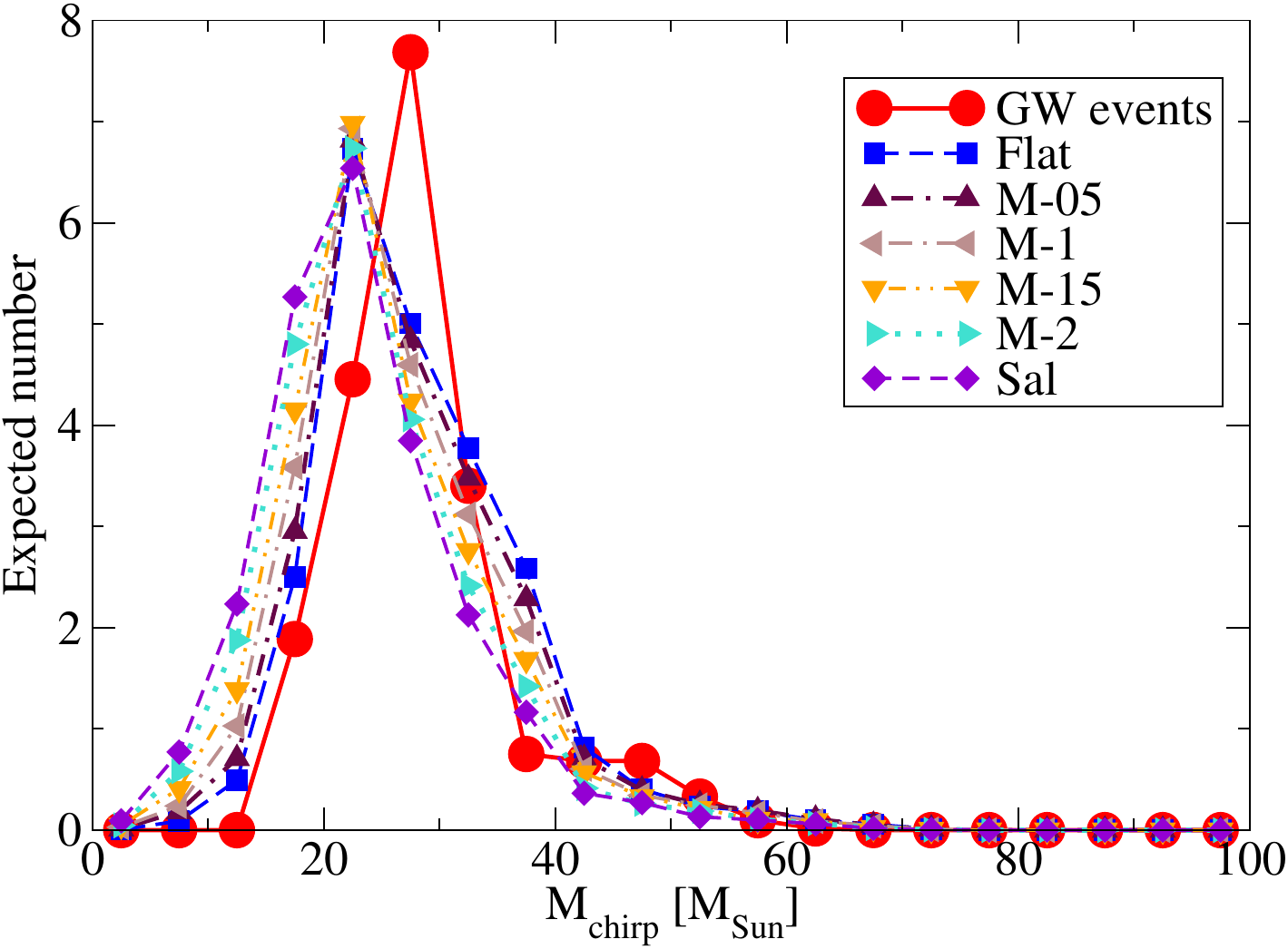}
  \end{center}
  \caption{Same figure of Fig.~\ref{fig:Mc_dist20} but for \citet{2021MNRAS.504L..28K}. The filled (red) circles show the observed 20 GW events, and the filled (blue) squares, filled (copper) triangles, filled (mustard) triangles, filled (khaki) triangles, filled (light blue) triangles, and filled (purple) diamonds denote the results of the Flat, M-05, M-1, M-15, M-2, and Sal models by assuming the total observable number is 20.}
  \label{fig:Mc_dist21}
\end{figure}

Figures~\ref{fig:Mc_dist20} and~\ref{fig:Mc_dist21} show the chirp-mass distribution of 20 BBH events with $\Mc > 18\,M_{\odot}$.
Figure \ref{fig:Mc_dist20} compares the GW result and models of \cite{2020MNRAS.498.3946K}.
The vertical axis shows the expected number of BBHs for a given range of $\Mc$ so that the total number should be 20. 
The filled (red) circles show the observed 20 GW events, and the filled (purple) triangles, filled (copper) triangles, filled (mustard) triangles, filled (khaki) triangles, filled (light blue) diamonds, filled (blue) squares, and  (green) asterisks denote the results of the al01, fiducial, FS1,FS2, K14, M100 and MT05 models by assuming the total observable number is 20. 
We can see that both the M100 and K14 models fit well with the observed one shown by the filled (red) circles. 

Figure \ref{fig:Mc_dist21} compares the GW result and models of \cite{2021MNRAS.504L..28K}.
The vertical axis shows the expected number of BBHs for a given range of $\Mc$ so that the total number should be 20. 
The filled (red) circles show the observed 20 GW events, and the filled (blue) squares, filled (copper) triangles, filled (mustard) triangles, filled (khaki) triangles, filled (light blue) triangles, and  filled (purple) diamonds denote the results of the Flat, M-05, M-1, M-15, M-2, Sal models by assuming the total observable number is 20. 
We can see that all models of \cite{2021MNRAS.504L..28K} have almost same shape and fit well with the observed one shown by the filled (red) circles. 

\begin{figure}
  \begin{center}
    \includegraphics[width=\hsize]{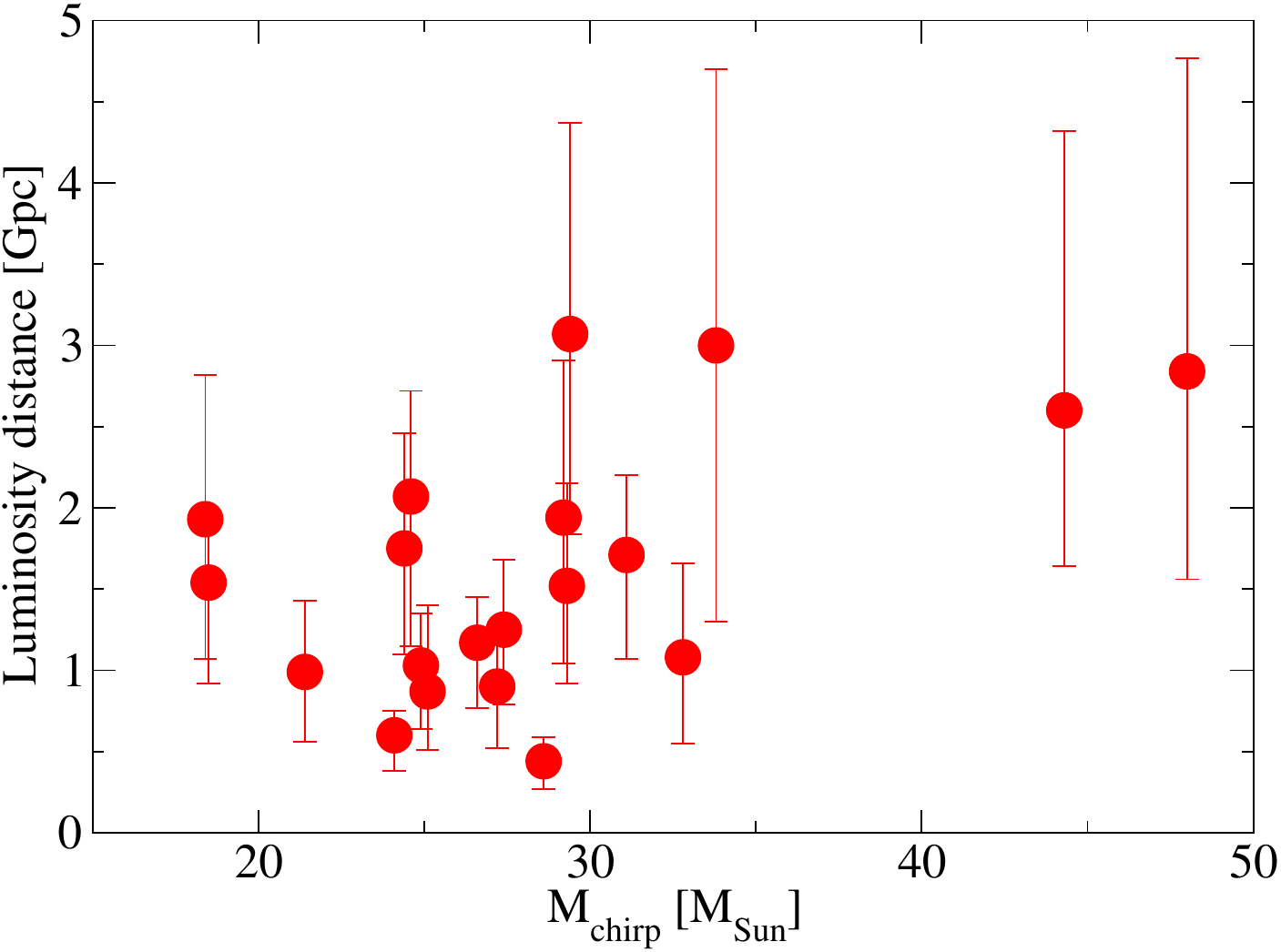}
  \end{center}
  \caption{Luminosity distance ($D_{\rm L}$) vs. chirp-mass of 20 BBH events with $\Mc > 18\,M_{\odot}$. Here, we present only the median of the chirp mass, and the median and 90\%-symmetric credible interval of the luminosity distance.}
  \label{fig:DL_dist}
\end{figure}

In Fig.~\ref{fig:Mc_dist20}, K14 and M100 models seem to match the data, while the other models are not suitable. In Fig.~\ref{fig:Mc_dist21}, all models appear to match the data. In other words, there are currently too many models that seem to be consistent with observations. This is because the current number of observations is too low. The only solution is to increase the number of observations by two or three times.
Figures~\ref{fig:Mc_dist20} and \ref{fig:Mc_dist21} suggest that, in order to match with observational results, it is crucial to have an initial mass range below $100\,\msun$ rather than focusing on the initial distribution, binary parameters, or the initial mass function (IMF). Note that \cite{2020MNRAS.498.3946K} did not account for the effects of pair-instability supernovae (PPISN). Considering PPISN, primary stars with initial masses above $100\,\msun$ lose mass due to PPISN, reducing their mass to around $30$--$50\,\msun$ \citep{Kinugawa2021}. This effect might lead to a reduction in the number of higher-mass black holes on the heavy side of the black hole mass distribution in models considering initial masses above $100\,\msun$, making them more consistent with observations.

Figure~\ref{fig:DL_dist} shows the luminosity distance ($D_{\rm L}$) distribution of 20 observed BBH events. 
We see that $D_{\rm L} \lesssim 3 \,{\rm Gpc}$ for $\Mc\lesssim 30\,\msun$ while $D_{\rm L} \gtrsim 1.5 \,{\rm Gpc}$ for $\Mc \gtrsim 30\,\msun$. But the errors are still large to say the above facts definitely.

\section{Discussion}

\begin{figure*}
  \begin{minipage}[b]{0.80\linewidth}
    \centering
    \includegraphics[width=\hsize]{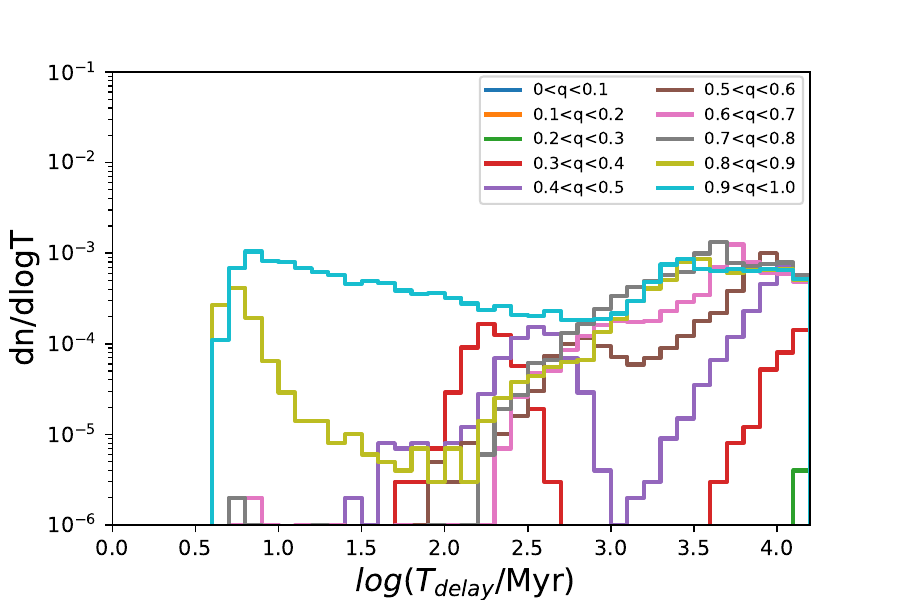}
    \subcaption[]{}
       \label{fig:QdistM100a}
    \end{minipage}
    \begin{minipage}[b]{0.80\linewidth}
    \centering
    \includegraphics[width=\hsize]{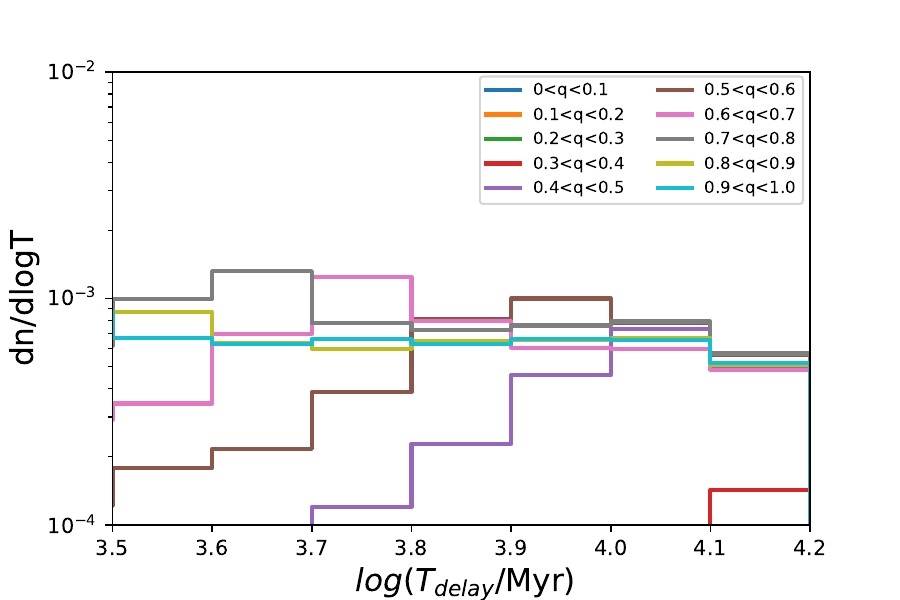}
    \subcaption[]{}
       \label{fig:QdistM100d}
  \end{minipage}
        \caption{Delay time ($T_{\rm delay}$) distributions for each mass ratio of merging Pop III BBHs in the M100 model. The distribution is normalized by the number of total Pop III binaries. (a) shows the delay time distributions of BBHs which merge within the Hubble time. (b) shows the delay time distributions of BBHs of which the merger time is more than 10$^{3.5}$ Myrs and less than the Hubble time.}\label{fig:QdistM100}
\end{figure*}

\begin{figure*}
   \begin{minipage}[b]{0.80\linewidth}
    \centering
    \includegraphics[width=\hsize]{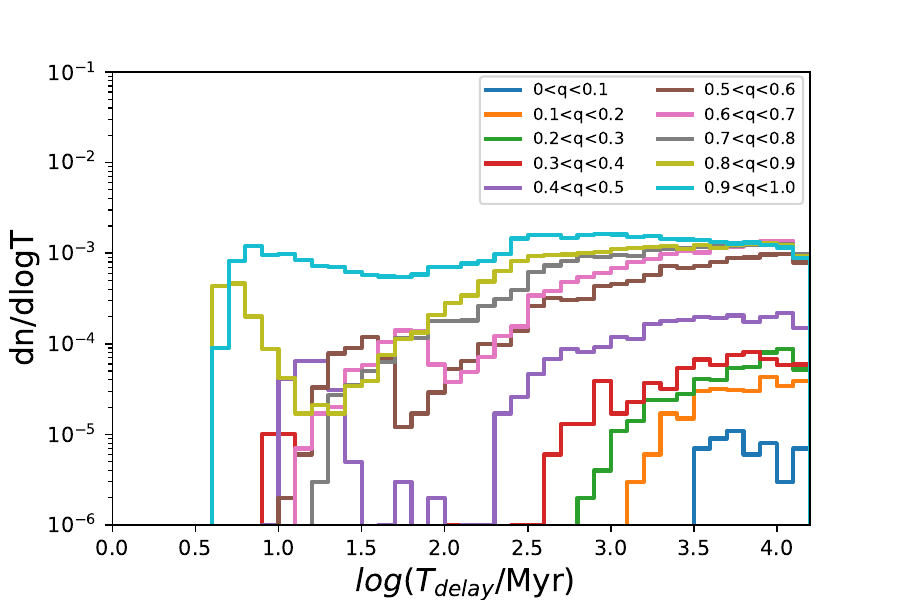}
        \subcaption[]{}
       \label{fig:QdistK14a}
    \end{minipage}
       \begin{minipage}[b]{0.80\linewidth}
    \centering
    \includegraphics[width=\hsize]{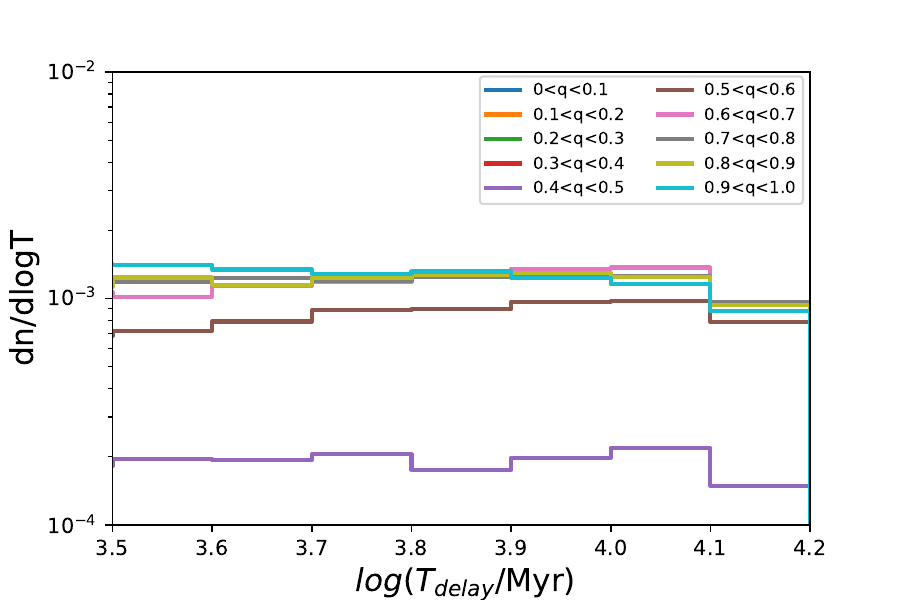}
        \subcaption[]{}
      \label{fig:QdistK14d}
    \end{minipage}
          \caption{Same figure of Fig.~\ref{fig:QdistM100} but for the K14 model.}
    \label{fig:QdistK14}
\end{figure*}

Figures~\ref{fig:QdistM100} and \ref{fig:QdistK14} show the delay time ($T_{\rm delay}$) distributions for each mass ratio of merging Pop III BBHs in the M100 and K14 models, respectively. 
We present the delay time distributions of Pop III BBHs which merge within the Hubble time in Figs. \ref{fig:QdistM100a} and \ref{fig:QdistK14a}, while the delay time distributions of Pop III BBHs of which the merger time is more than 10$^{3.5}$ Myrs and less than the Hubble time in Figs. \ref{fig:QdistM100d} and \ref{fig:QdistK14d}.
These distributions are normalized by the number of total Pop III binaries.

It is found in Figs. \ref{fig:QdistM100a} and \ref{fig:QdistK14a} that the nearly equal-mass BBH mergers predominate in a very short delay time region ($T_{\rm delay} \lesssim100$ Myrs).
In other words, the nearly equal-mass Pop III BBH mergers predominate in the very early universe ($z\gtrsim10$) since the Pop III stars are born and died at high redshift ($z\gtrsim10$).
These BBH mergers at the high redshift can be observed future GW observatory such as the Einstein telescope (ET)~\citep{ET}, the Cosmic Explorer (CE)~\citep{CE}, and DECIGO~\citep{Seto:2001qf,Nakamura_2016}.

The formation channel is the reason why the equal mass Pop III BBHs merge at the high redshift.
In our previous paper \citep{2020MNRAS.498.3946K}, we classified the Pop III BBH formation channel into the 5 channels such as NoCE, 1CE$_P$, 1CE$_S$, 1CE$_D$, and 2CE.
NoCE means that Pop III BBHs evolve not via a common envelope phase. 1CE$_P$, 1CE$_S$, and 1CE$_D$ are Pop III BBHs evolved via one common envelope phase caused by the primary giant star, the secondary giant star, or double giant stars, respectively. 2CE is the Pop III BBHs experienced more than two common envelope phases.
Figure 5 of~\cite{2020MNRAS.498.3946K} shows that almost all Pop III BBHs merging with very short delay time ($T_{\rm delay} \lesssim100$ Myrs) evolved via the 1CE$_D$ channel.

\begin{figure}
  \begin{center}
    \includegraphics[width=\hsize]{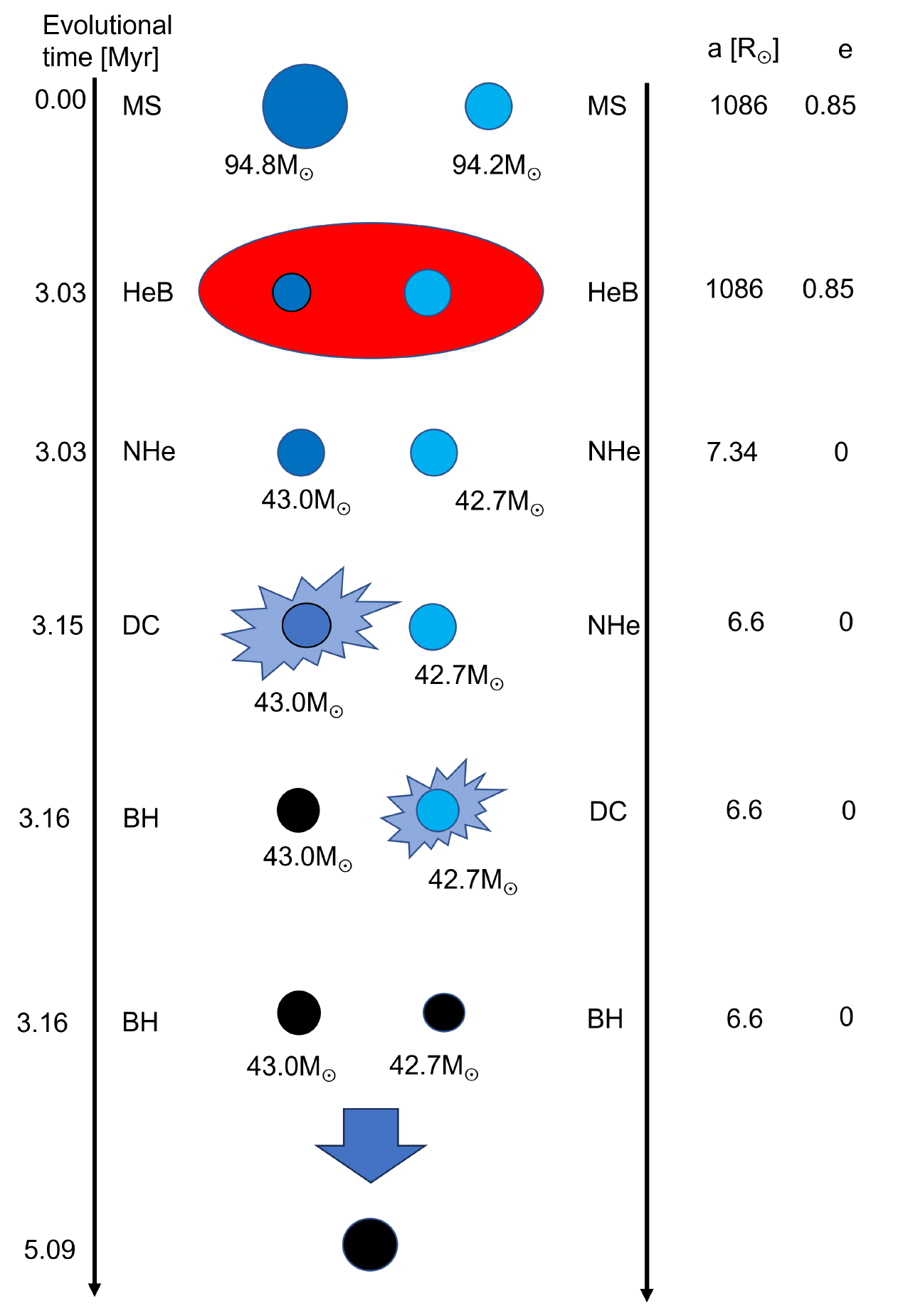}
  \end{center}
  \caption{Example of Pop III binary evolution via 1CE$_D$ channel. Here, $a$ and $e$ denote the separation and eccentricity, respectively. MS, HeB, NHe, DC and BH mean main-sequence phase, helium-burning phase, naked helium star, degenerate core, and black hole, respectively.}
  \label{fig:equalmassBBH}
\end{figure}

Figure \ref{fig:equalmassBBH} shows an example of 1CE$_D$ channel.
Pop III binaries of which initial masses are nearly equal tend to evolve via the 1CE$_D$ channel.
If the initial masses of a binary are similar, the timescale of evolution is similar as well. Therefore, they both become giant stars at almost the same time and shed their envelopes simultaneously through a double common envelope process. 
A significant amount of orbital energy is also lost by shedding the outer envelopes of both stars.
After that, a very close binary He star tends to remain, and they can become a close BBH which merges within 100 Myrs.

\begin{figure}
  \begin{center}
    \includegraphics[width=\hsize]{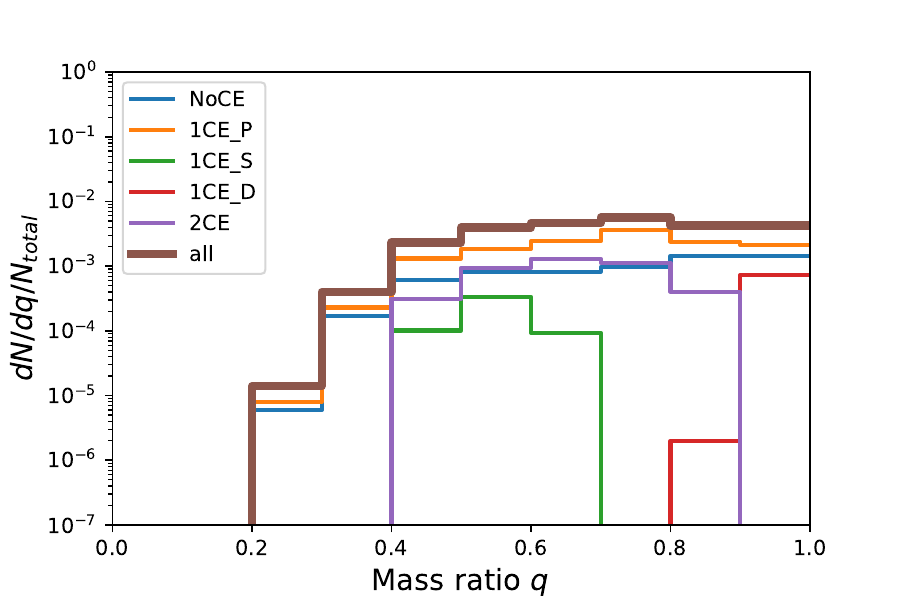}
  \end{center}
  \caption{Mass ratio distributions of merging Pop III BBHs of which delay time is more than $10^{3.5}$ Myrs and less than the Hubble time
for M100 model.}
  \label{fig:longmergertime_M100}
\end{figure}

\begin{figure}
  \begin{center}
    \includegraphics[width=\hsize]{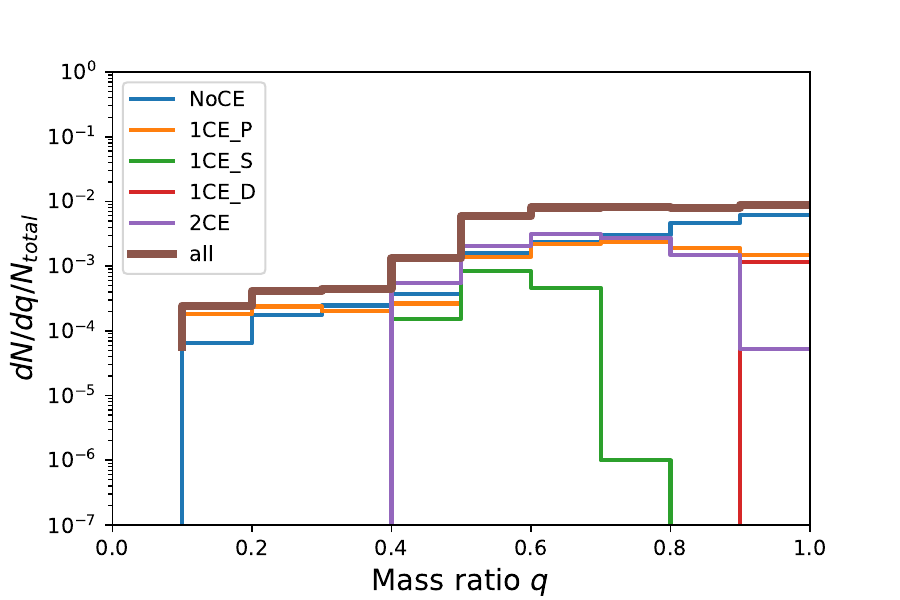}
  \end{center}
  \caption{Same figure of Fig.~\ref{fig:longmergertime_M100} but for the K14 model.}
  \label{fig:longmergertime_K14}
\end{figure}

Next, we focus on Figs. \ref{fig:QdistM100d} and \ref{fig:QdistK14d} where Pop III BBHs with a long delay time are presented for each mass ratio.
Figures \ref{fig:longmergertime_M100} and \ref{fig:longmergertime_K14} show the mass ratio distributions of merging Pop III BBHs of which delay time is more than 10$^{3.5}$ Myrs and less than the Hubble time for M100 and K14 models, respectively.
These BBHs can be detected within the detection range of LVK collaboration.

In the M100 model, the main contribution to the mass ratio distribution is the 1CE$_P$ channel. 
Subdominant ones are NoCE and 2CE channels. 
These channels make BBHs with various mass ratios from 0.4 to 1, unlike the 1CE$_D$ channel.
In the K14 model, the main contribution is the NoCE channel. 
Subdominant ones are 1CE$_P$ and 2CE channels. 
This model also has various mass ratio BBH mergers with a long delay time like the M100 model.

{\section{Summary and Conclusion}}

{In this paper,} we focus on gravitational-wave events of binary black-hole mergers up to the third observing run with the minimum false alarm rate smaller than $10^{-5}\,{\rm yr}^{-1}$. 
These events tell us that the mass ratio of two black holes follows $m_2/m_1=0.723$ with the chance probability of 0.00301\% for $\Mc > 18\,M_{\odot}$ where $\Mc$ is called the chirp mass of binary. We show that the relation of $m_2/m_1=0.723$ is consistent with the binaries originated from Pop III stars which are the first stars in the universe. 

On the other hand, it is found for $\Mc < 18 M_{\odot}$ that the mass ratio follows $m_2/m_1=0.601$ with the chance probability of 0.117\% if we ignore GW190412 with $m_2/m_1\sim 0.32$. This suggests a different origin from that for $\Mc > 18 M_{\odot}$.
For $\Mc< 18\msun$, there is one with significantly different $m_2/m_1$, and the correspondence to Pop I and Pop II is not clear. To state anything conclusively, the number of observed events needs to increase by several times. Also, in future observations, events with higher redshifts will become visible.

Furthermore, according to Figs.~\ref{fig:QdistM100}, \ref{fig:QdistK14}, \ref{fig:longmergertime_M100} and \ref{fig:longmergertime_K14}, it is expected that observations at high redshift, unlike observations at low redshift, would exhibit a higher proportion of equal-mass BBH mergers.
A comparison between the low-redshift results from the current ground-based GW observations and the high-redshift results from future observations such as ET, CE and DECIGO enables us to check the Pop III origin model.

\vspace{10mm}

\section*{Acknowledgment}

T. K. acknowledges support from JSPS KAKENHI Grant Numbers JP21K13915 and JP22K03630.
H. N. acknowledges support from JSPS KAKENHI Grant Numbers JP21H01082, JP21K03582 and JP23K03432,
and also would like to thank to Y. W. for her hospitality.

\section*{Data Availability}

Results will be shared on reasonable request to the corresponding author.

\bibliographystyle{mnras}

\bibliography{ref}

\end{document}